\documentclass[10pt,conference]{IEEEtran}
\IEEEoverridecommandlockouts
\usepackage{subcaption}
\usepackage{cite}
\usepackage{amsmath,amssymb,amsfonts}
\PassOptionsToPackage{hyphens}{url}
\usepackage[hidelinks]{hyperref}
\usepackage{graphicx}
\usepackage{textcomp}
\usepackage{xcolor}
\usepackage{multirow}
\usepackage{booktabs}
\usepackage{dsfont}
\usepackage{enumitem}
\usepackage{algorithm}
\usepackage{algpseudocode}
\def\BibTeX{{\rm B\kern-.05em{\sc i\kern-.025em b}\kern-.08em
    T\kern-.1667em\lower.7ex\hbox{E}\kern-.125emX}}
\begin{document}

\title{Dual Explanations via Subgraph Matching for Malware Detection}

\author{\IEEEauthorblockN{Hossein Shokouhinejad, Roozbeh Razavi-Far, Griffin Higgins, Ali A. Ghorbani}
\IEEEauthorblockA{\textit{Canadian Institute for Cybersecurity (CIC), University of New Brunswick}, Fredericton, NB, Canada\\
Email: \{hossein.shokouhinejad, roozbeh.razavi-far, griffin.higgins, ghorbani\}@unb.ca}}

\maketitle

\begin{abstract}
Interpretable malware detection is crucial for understanding harmful behaviors and building trust in automated security systems. Traditional explainable methods for Graph Neural Networks (GNNs) often highlight important regions within a graph but fail to associate them with known benign or malicious behavioral patterns. This limitation reduces their utility in security contexts, where alignment with verified prototypes is essential. In this work, we introduce a novel dual prototype-driven explainable framework that interprets GNN-based malware detection decisions. This dual explainable framework integrates a base explainer (a state-of-the-art explainer) with a novel second-level explainer which is designed by subgraph matching technique, called SubMatch explainer. The proposed explainer assigns interpretable scores to nodes based on their association with matched subgraphs, offering a fine-grained distinction between benign and malicious regions. This prototype-guided scoring mechanism enables more interpretable, behavior-aligned explanations. Experimental results demonstrate that our method preserves high detection performance while significantly improving interpretability in malware analysis.

\end{abstract}


\section{Introduction}

Graph Neural Networks (GNNs) have emerged as powerful tools for learning from structured data and have shown strong performance across various domains including social networks, chemistry, and cybersecurity. In the context of malware detection, GNNs are particularly well-suited due to their ability to model complex program structures, such as control flow graphs (CFGs), which capture the execution behavior of binary programs. By leveraging the topological information and semantic relationships within these graphs, GNN-based methods can identify subtle malicious patterns that are often missed by traditional techniques. Recent studies have demonstrated the effectiveness of applying GNNs to malware classification, graph-based anomaly detection, and behavioral analysis of executable files \cite{01,02,03,04,05,06,07,08,09,10,11,12,13,14,ZHEN2025110524}. Specifically, dynamically extracted CFGs from PE files provide rich behavioral signals, allowing GNNs to learn patterns that correspond to malicious intent or benign operations \cite{Our_Survey}.

Despite the impressive performance of GNNs in malware detection and other graph-based tasks, their black-box nature poses significant challenges in understanding and trusting their decisions~\cite{Hesam}. To address this, several GNN-specific explainability techniques have been proposed. GNNExplainer is one of the earliest and most widely used methods, which identifies a compact subgraph and a subset of node features that are most influential to the model's prediction for a given instance \cite{10.5555/3454287.3455116}. PGExplainer takes a probabilistic approach by learning a parameterized distribution over possible explanations across multiple samples, improving generalization and scalability \cite{PG}. Frameworks like Captum, designed for PyTorch models, also offer GNN support and provide model-agnostic interpretability methods such as integrated gradients and saliency maps \cite{captum}. More recently, SubgraphX has introduced a Monte Carlo tree search-based strategy to explore subgraphs and compute Shapley values, offering both accuracy and theoretical guarantees for the importance of the extracted subgraphs \cite{subgraphx}. While these methods provide useful insights into which parts of a graph contribute most to a GNN’s prediction, they typically operate on a per-sample basis and lack the ability to link explanations to known malicious or benign behavior patterns. This limits their utility in malware analysis, where interpretability must often go beyond local graph regions to include behavioral context and prototype associations.

Subgraph matching has gained attention as a means of discovering repeated behavioral patterns in malware samples. It enables structural comparisons between graphs, making it a useful technique for identifying known malicious or benign behaviors embedded within complex software structures. While subgraph matching has been used in various domains, its potential for interpretable malware detection remains largely underexplored. A few recent studies have incorporated subgraph analysis into malware classification tasks, though with objectives and designs different from ours. For instance, Fan et al. introduced FalDroid, a system that uses frequent subgraph analysis to classify Android malware into families and select representative samples for efficient investigation. Their focus is on familial clustering rather than fine-grained interpretability within individual samples \cite{8293854}. Similarly, the SNDGCN framework employs subgraph networks (SGNs) to improve adversarial robustness in Android malware detection, but it does not link subgraph structure to interpretable detection outcomes \cite{LU2024123922}. Another approach, MARD, proposed by Alam et al., applies a matching strategy based on Annotated Control Flow Graphs (ACFGs) to detect metamorphic malware in real-time, targeting fast and obfuscation-resilient detection rather than explainability \cite{ALAM2015212}. These efforts show the viability of subgraph-based techniques for malware analysis, yet none of them attempt to distinguish malicious and benign behaviors within the same graph or provide interpretable, node-level insights into model decisions. This leaves a critical gap that our method aims to fill.

While existing GNN explainers provide local explanations by highlighting important nodes or edges, they do not offer insight into how these components relate to known patterns of malicious or benign behavior. This limitation reduces their usefulness in malware detection tasks, where understanding both the origin and nature of suspicious behaviors is essential. To address this, we propose a dual interpretability framework that integrates subgraph matching with GNN explainability to provide behavior-aware, prototype-based interpretations. The key contributions of this paper are as follows:
The key contributions of this paper are as follows:
\begin{itemize}[left=0pt]
    \item We propose a novel dual prototype-driven explainable framework that integrates a base GNN explainer with a second-level explainer based on subgraph matching, enabling behavior-aligned interpretation for GNN-based malware detection.
    \item We design the SubMatch explainer as the second-level component of the framework, which performs subgraph matching between the target malicious graph and verified subgraphs to assign interpretable node scores based on their structural associations.
    \item Our method enables fine-grained interpretation by distinguishing between malicious and benign regions within the same graph, an ability that existing explainers lack.
\end{itemize}

The rest of this paper is organized as follows. Section~\ref{sec:preliminaries} provides the necessary preliminaries and background information relevant to graph-based malware detection and subgraph matching. Section~\ref{sec:methodology} describes our proposed novel dual explainer methodology, including the design of the SubMatch explainer. Section~\ref{sec:experiments} presents the experimental setup and analyzes the obtained results to demonstrate the effectiveness of the proposed framework. Finally, Section~\ref{sec:conclusion} concludes the paper.

\section{Preliminaries}
This section provides essential background information to support the proposed framework. We begin by introducing the foundational principles of GNNs. Next, we discuss various GNN explainers that offer interpretability by highlighting influential components within the graph structure. Finally, we present subgraph matching techniques, which enable the alignment of extracted graph patterns with known benign or malicious prototypes.

\label{sec:preliminaries}
\subsection{Graph Neural Networks}
GNNs are a class of deep learning models designed to operate on graph-structured data. In contrast to conventional neural networks that assume a fixed input structure, GNNs can directly model the irregular and non-Euclidean structure of graphs, making them highly effective for tasks involving relational data such as social networks, knowledge graphs, and program analysis.

A GNN learns a representation for each node by iteratively aggregating and transforming information from its neighboring nodes over multiple layers~\cite{NCP}. The goal is to capture both local and global structural information of the graph. The representation of node \( i \) at the \( l \)-th layer is typically computed using the following general rule:
\begin{equation}
h^{(l)}_i = \text{UPDATE}\left(h^{(l-1)}_i, \text{AGG}\left(\{h^{(l-1)}_j \mid j \in \mathcal{N}(i)\}\right)\right)
\end{equation}
where \( h^{(l)}_i \) is the embedding of node \( i \) at layer \( l \), \( h^{(0)}_i \) is the initial feature vector of node \( i \), and \( \mathcal{N}(i) \) denotes the set of its neighbors. The function \( \text{AGG}(\cdot) \) aggregates information from the neighbors, and \( \text{UPDATE}(\cdot) \) updates the node's representation based on its previous state and the aggregated messages.

In the context of malware detection, GNNs are particularly well-suited for analyzing CFGs extracted from executable files. These graphs represent the execution logic of programs, where nodes correspond to code blocks (e.g., basic blocks or functions) and edges represent control flow transitions. By learning node embeddings that encode both syntactic and structural properties of the code, a GNN can capture high-level behavioral patterns of benign and malicious software.

Through message passing across CFG, the GNN model can generalize from known malicious patterns to detect obfuscated or structurally altered variants. As a result, GNNs have become a powerful backbone for graph-based malware detection systems, offering robustness against code-level transformations that commonly evade traditional signature-based methods.
\subsection{GNN Explainers}

GNN Explainers are interpretability techniques designed to uncover which components of an input graph contribute most significantly to a GNN's prediction. By assigning importance scores to elements of the graph, these methods provide insight into the decision-making process of the model and help to identify the substructures that drive its output.

Most GNN explainers focus on assigning scores to edges, as edge connectivity governs the message-passing mechanism that underlies GNNs. These edge importance scores indicate how much each connection influences the flow of information and ultimately affects the model’s prediction. Some explainers, such as GNNExplainer, also provide node-level scores to highlight the relevance of specific nodes in the prediction process.

Given an input graph \( G = (V, E) \), the explainer produces a set of edge importance scores:
\begin{equation}
\mathbf{S}_E = \{\alpha_e : e \in E\}
\end{equation}
where \( \alpha_e\) reflects the contribution of edge \( e \) to the model’s output. Higher values of \( \alpha_e \) indicate greater influence.

The explanation can be represented as the original graph annotated with these scores:
\begin{equation}
G' = (V, E, \mathbf{S}_E)
\end{equation}
where \( G' \) maintains the original topology of \( G \) but includes additional information in the form of edge importance weights.

Different explainer models use various strategies to generate these scores. For example, optimization-based approaches like GNNExplainer aim to find a compact subgraph that preserves the model’s output. Probabilistic models such as PGExplainer learn distributions over important edges, while search-based methods like SubgraphX explore the space of possible subgraphs using techniques like Monte Carlo Tree Search. Despite their differences, all these methods aim to enhance the transparency and trustworthiness of GNN predictions by revealing the most influential structural components.

\subsection{Subgraph Matching}

Subgraph matching is a fundamental operation in graph analysis that involves identifying occurrences of a smaller query graph within a larger target graph. It plays a central role in numerous domains such as bioinformatics, computer vision, social network analysis, and, notably, cybersecurity and malware detection. In this context, CFGs representing program behavior can be analyzed for known malicious patterns through subgraph matching techniques.

There are two primary forms of subgraph matching: subgraph isomorphism and subgraph monomorphism.

\begin{itemize}[left=0pt]
    \item \textbf{Subgraph Isomorphism} requires an exact, structure-preserving, and label-consistent mapping from the query graph to a subgraph of the target graph. Formally, given a query graph \( Q = (V_Q, E_Q) \) and a target graph \( T = (V_T, E_T) \), a function \( f: V_Q \to V_T \) is a subgraph isomorphism if:
    \begin{enumerate}
        \item \( f \) is injective.
        \item For every \( (u, v) \in E_Q \), \( (f(u), f(v)) \in E_T \).
        \item Vertex and edge labels are preserved under \( f \).
    \end{enumerate}

    \item \textbf{Subgraph Monomorphism} relaxes the strict requirements of isomorphism. While it still requires injectivity and structural preservation of query edges, it allows the presence of additional edges in the target graph that are not part of the query. This makes monomorphism suitable for scenarios where additional context exists around a matching structure, such as benign extensions around malicious code blocks.
\end{itemize}

One of the most widely used algorithms for subgraph matching is VF2. It efficiently searches for valid node mappings by exploring the state space through depth-first traversal and applying pruning rules to eliminate infeasible candidates early. At each state, VF2 incrementally builds a partial mapping \( M \) between the query and target graphs. Candidate pairs \( (u, v) \), with \( u \in V_Q \) and \( v \in V_T \), are evaluated under structural and semantic feasibility rules to ensure consistent mappings. Notably, VF2 supports both subgraph isomorphism and subgraph monomorphism matching types, making it flexible for different application needs. Additionally, the algorithm can incorporate vertex and edge attributes into its feasibility checks, enabling more fine-grained and context-aware matching.
\section{Proposed Methodology}
\label{sec:methodology}
While several GNN explanation methods have been developed, most provide low-level attributions in the form of node or edge importance scores. These scores, although useful for general graph analysis, often lack the semantic clarity needed for malware investigation. They do not explicitly relate prediction outcomes to meaningful behavioral patterns in the software, making the explanations difficult to interpret in practice.

To address these limitations, we propose a novel dual explainer within malware detection framework that incorporates subgraph-level reasoning into the interpretability process. In the training phase, CFGs are dynamically extracted from PE files, and node features are embedded using an autoencoder. These embeddings are used to train a GNN model for classification. A GNN explainer is then applied to correctly labeled samples to extract important subgraphs. These subgraphs (explanations) are verified by re-evaluating them through the pre-trained GNN model, and only those that retain their labels are stored in a verified query box as trusted benign or malicious prototypes.

In the test phase, each sample is first classified using the pre-trained GNN model. Only the samples predicted as malicious, referred to as target samples, proceed to the interpretation step. These target samples are matched against the verified subgraphs using a subgraph matching algorithm. The SubMatch Explainer then analyzes the associations between regions of the target graph and the verified subgraphs, assigning interpretable scores that reflect their similarity to known benign or malicious patterns. This process results in behavior-grounded and context-aware explanations that enhance the transparency and reliability of GNN-based malware detection.

\begin{figure*}[h]
    \centering
    \includegraphics[width=\textwidth]{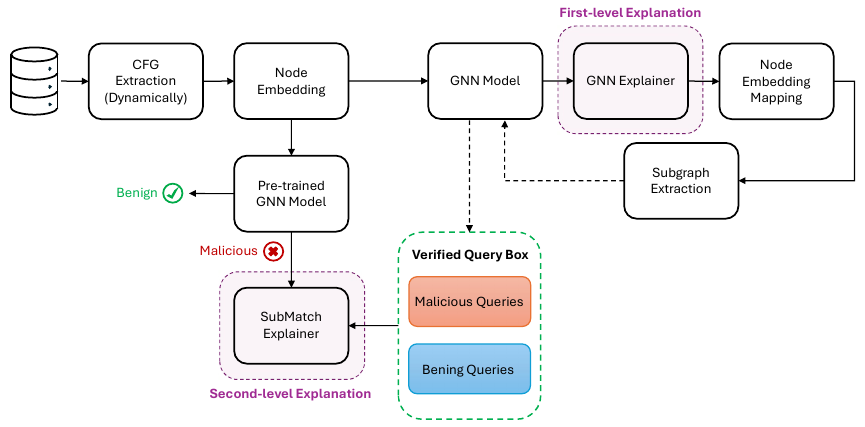}
    \caption{The proposed malware detection framework for dual explanation of malicious samples.}
    \label{fig:framework}
\end{figure*}

In the remainder of this section, we present the proposed methodology in detail. The proposed framework is composed of multiple stages, as illustrated in Figure~\ref{fig:framework}.

\subsection{Dynamic CFG Extraction}
The process begins with dynamic extraction of CFGs from program execution. Let \( \mathcal{F} \) denote the set of programs under analysis, and each program \( f \in \mathcal{F} \) is executed in a controlled environment to generate its corresponding CFG \( G_{f} = (V_{f}, E_{f}) \), where:
\begin{itemize}
    \item \( V_{f} \) is the set of nodes representing basic blocks or instructions.
    \item \( E_{f} \subseteq V_{f} \times V_{f} \) is the set of directed edges representing control flow transitions.
\end{itemize}
The CFG captures the structural and behavioral properties of the program, enabling the subsequent analysis.

\subsection{Node Feature Embedding}

Each node in a CFG may contain a variety of raw information, such as assembly instructions, system calls, or code block metadata. However, not all of this information is directly suitable for use in GNNs, which require fixed-size, numerical input. The goal of this stage is to extract meaningful attributes from each node and transform them into low-dimensional vector representations that preserve semantic information relevant for classification. The final output of this stage is a set of node embeddings that capture the functional characteristics of the CFG nodes in a format suitable for downstream GNN-based analysis.

Once the embeddings are generated, the CFGs are divided into training and test datasets. The training set is used for training the GNN model, applying the GNN explainer, and extracting subgraphs, while the test set is subjected to second-level explanation (SubMatch Explainer).

\subsection{GNN-Based Classification}

In this stage, a GNN model is trained on the CFGs to distinguish between benign and malicious samples. The GNN learns to capture both local and global structural patterns by aggregating node information across the graph, ultimately producing a graph-level representation used for classification. Once the model is trained, we evaluate its predictions on the training set and select only those samples that are correctly classified by the GNN. These correctly predicted samples are then forwarded to the next stage, where they serve as inputs to the explanation module for subgraph extraction.

\subsection{GNN-based Explainer and Node Embedding Mapping}
For samples correctly classified as benign or malicious, we employ a graph explanation module designed to identify the most critical nodes and edges that contribute to the predictions made by the GNN. This module provides insights into the model’s decision-making process by highlighting specific substructures within the graph that are most influential in determining the classification outcome. The output of the explainer includes the original graph structure, augmented with importance weights assigned to edges, nodes, or both. These weights reflect the relative contribution of each component to the final prediction.

The output graph from GNN explainers typically lacks the original node feature embeddings. To support downstream tasks like subgraph matching, we reattach the original embeddings to their corresponding nodes in a process called Node Embedding Mapping. This step ensures that the semantic information captured during preprocessing is preserved, enabling further analysis that depends on both structural and contextual node attributes.
\begin{figure*}
    \centering
    \includegraphics[width=\linewidth]{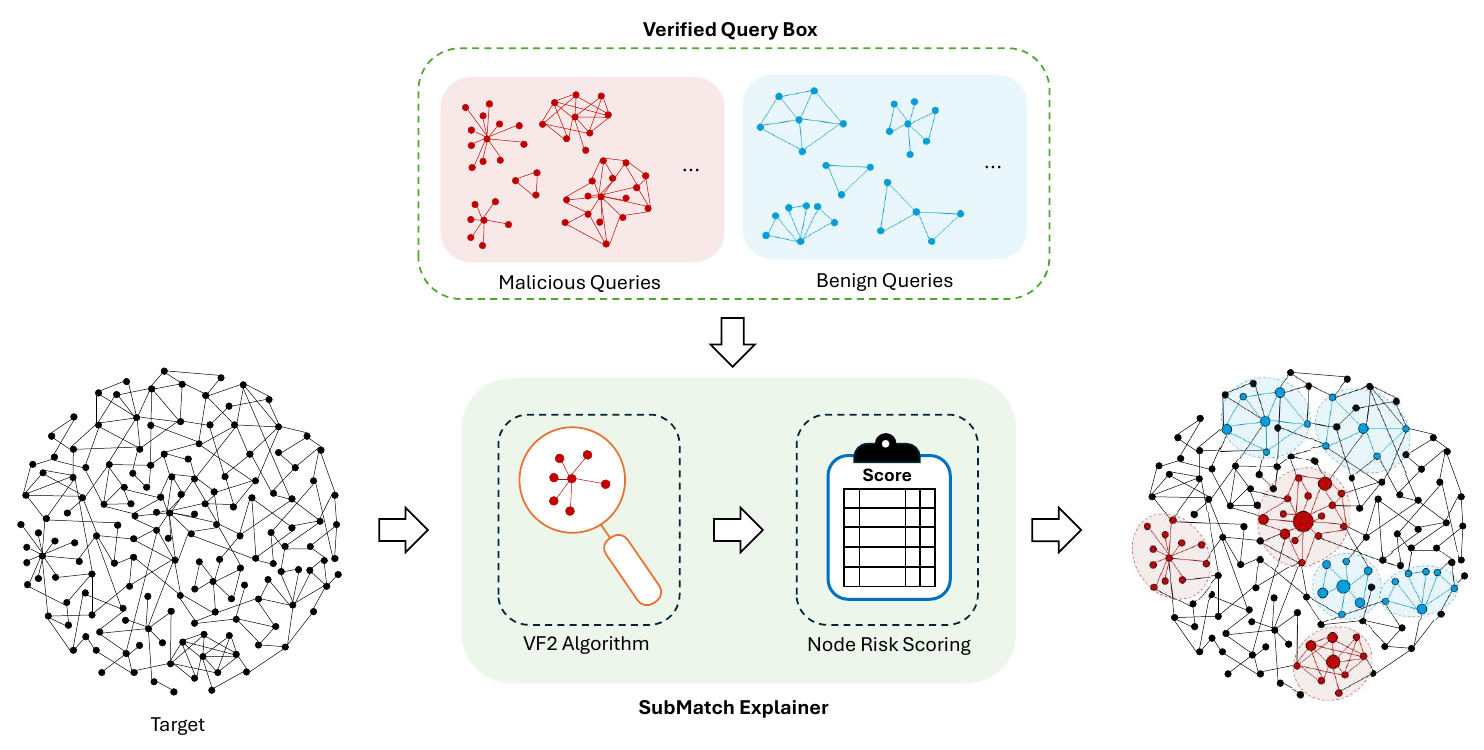}
    \caption{Subgraph matching visualization using SubMatch explainer (second-level for dual explanation). Node color indicates match type (red: malicious, blue: benign), and node size reflects the match frequency.}
    \label{fig:SubMatch_Explainer}
\end{figure*}

\subsection{Subgraph Extraction}

Our proposed framework proceeds with subgraph extraction and verification. The process begins with the extraction of subgraphs from the weighted graph output of the explainability module. To extract the most informative and structurally meaningful part of the graph, we propose Greedy Edge-wise Composition (GEC)~\cite{Dynamic}, an alternative subgraph extraction technique. The primary objective of GEC is to construct a strongly connected subgraph that maximizes the total edge weight, thereby retaining the most critical structural information identified by the explainability module.

The procedure begins by identifying the edge with the greatest importance score from the weighted graph produced by the GNN explanation module. Let this initial edge be denoted as $e_{\text{max}} = \displaystyle\arg\max_{\substack{e \in E}} \alpha_e$ represents the weight of edge $e$. The corresponding nodes of $e_{\text{max}}$ are added to the selected node set, and the edge is added to the selected edge set.

Subsequently, GEC iteratively adds one edge at a time by selecting the highest-weight edge that connects to the current set of selected nodes:
\[
e_{\text{next}} = \arg\max_{\substack{e \in E}} \alpha_e, \quad e \text{ connects to } V_{\text{selected}}.
\]
This greedy selection continues until a predefined number of edges (or nodes) is reached, ensuring both high cumulative edge weight and strong connectivity within the resulting subgraph.

Let $k$ be the desired number of edges to include. The final extracted subgraph is:
\[
G_{\text{extracted}} = (V_{\text{selected}}, E_{\text{selected}}), \quad \text{where } |E_{\text{selected}}| = k.
\]

By prioritizing both edge weight and graph connectivity, GEC produces subgraphs that more faithfully reflect the underlying structure of the original graph. This leads to more reliable explanations and a better understanding of the model’s decision-making behavior.

Once subgraphs are extracted, they undergo a verification step to ensure consistency with the original classification labels of their corresponding samples. This is achieved by feeding each subgraph back into the pre-trained GNN model and assessing whether it produces the same label as the original full graph \( G \). Subgraphs that satisfy this condition, i.e., \( y_{\text{subgraph}} = y_{\text{original}} \), are considered reliable representations of the original samples. These validated subgraphs are subsequently stored in the query box, which serves as a curated repository of high-confidence subgraphs used in the downstream matching process.

\subsection{SubMatch Explainer}
The SubMatch explainer operates as a second-level interpretability mechanism, activated during the test phase to provide behavior-aligned explanations for samples flagged as malicious by the GNN classifier. Once a target graph \( \mathcal{T} \) is identified, SubMatch initiates a two-stage process designed to connect structural patterns in the graph to known malicious or benign behaviors.

Specifically, SubMatch consists of two core components. The first component, Subgraph Matching, identifies regions of structural or semantic similarity between the target graph and a repository of verified subgraphs, including both benign and malicious prototypes. The second component, node risk scoring, quantifies the contribution of each node to potentially harmful behavior by assigning interpretable scores based on its association with these matched subgraphs.

\subsubsection{Subgraph Matching}
The Subgraph matching procedure compares the target graph with the verified subgraphs stored in the query box, defined as \( \mathcal{B} = \mathcal{B}_{\text{mal}} \cup \mathcal{B}_{\text{ben}} \). Here, \( \mathcal{B}_{\text{mal}} = \{ \mathcal{Q}_{m1}, \mathcal{Q}_{m2}, \ldots, \mathcal{Q}_{m_k} \} \) represents the set of verified malicious subgraphs, and \( \mathcal{B}_{\text{ben}} = \{ \mathcal{Q}_{b1}, \mathcal{Q}_{b2}, \ldots, \mathcal{Q}_{b_\ell} \} \) denotes the set of verified benign subgraphs. The objective of this matching step is to identify regions within the target graph that are structurally and semantically similar to subgraphs correspond to either known malicious or benign behavior.

To accomplish this, the VF2 algorithm is employed in conjunction with two distinct matching strategies. The first strategy, exact structural and attribute matching, enforces strict alignment between both the graph topology and node attributes. This ensures a high degree of precision by requiring that the structural connections and node features in the target graph exactly match those in the query subgraphs.

The second strategy, exact structural matching with approximate attribute matching, relaxes the attribute-level constraint while maintaining topological fidelity. In this approach, node embeddings are compared using cosine similarity to allow for semantic alignment despite minor variations in raw feature values. Let \( \phi: V \to \mathbb{R}^d \) denote the embedding function that maps each node to a \( d \)-dimensional vector. For a node \( v \in V_{\mathcal{Q}_i} \), a match is established with node \( f(v) \in V_{\mathcal{T}} \) if their embeddings satisfy the condition:
\[
\cos\left(\phi_{\mathcal{Q}_i}(v), \phi_{\mathcal{T}}(f(v))\right) \geq \delta,
\]
where \( \delta \in [0,1] \) is a predefined similarity threshold. This relaxed strategy enhances robustness by enabling the detection of functionally similar subgraphs that may not share identical attribute representations, thereby increasing tolerance to noise and obfuscation in node-level features.

For each verified subgraph \( \mathcal{Q}_i \in \mathcal{B} \), the matching process identifies substructures in the target graph \( \mathcal{T} \) that satisfy the corresponding matching criteria. The selection between the two approaches depends on the desired balance between precision and generalization. Exact structural and attribute matching enforces strict alignment of both the graph topology and node attributes, leading to high-confidence matches. In contrast, approximate attribute matching relaxes the constraints on node features by allowing minor variations in the embedding space, while preserving the exact structural layout, offering greater flexibility in identifying semantically similar patterns.

\subsubsection{Node Risk Scoring}
In our proposed SubMatch explainer, we introduce a scoring mechanism to interpret malware explanation results by analyzing the associations of each node in the target graph with subgraphs in a verified query box. The SubMatch explainer leverages subgraph matching to meticulously identify which graph components contribute to malicious or benign behaviors, offering a level of granularity that surpasses traditional GNN explainers. Its functionality revolves around assigning scores to nodes based on their matches with either malicious or benign subgraphs, emphasizing interpretability and precision.

Each node in the target graph is evaluated as follows:
\begin{itemize}[left=0pt]
    \item Nodes matching malicious subgraphs are scored \textbf{+1} for each match, and their score increases cumulatively with additional matches, reflecting a stronger association with malicious patterns.
    \item Nodes matching benign subgraphs are scored \textbf{-1} for each match, with their score decreasing cumulatively for multiple matches, indicating a stronger association with benign behaviors.
    \item Nodes that do not match any subgraph remain unscored, resulting in a score of \textbf{0}, signifying neutrality with no relation to either malicious or benign patterns.
\end{itemize}

The score for each node \( v \) in the target graph is computed as:
\[
S(v) =
\begin{cases}
-\sum\limits_{\mathcal{Q}_b \in \mathcal{B}_{\text{ben}}} \mathbb{I}(v \in V_{\mathcal{Q}_b}) \\
\sum\limits_{\mathcal{Q}_m \in \mathcal{B}_{\text{mal}}} \mathbb{I}(v \in V_{\mathcal{Q}_m}) \\
0, \quad \text{otherwise}.
\end{cases}
\]
Here, for each subgraph \( \mathcal{Q}_m \in \mathcal{B}_{\text{mal}} \) and \( \mathcal{Q}_b \in \mathcal{B}_{\text{ben}} \), \( V_{\mathcal{Q}_m} \) and \( V_{\mathcal{Q}_b} \) denote their corresponding node sets. The indicator function \( \mathbb{I}(\cdot) \) returns 1 if the condition holds and 0 otherwise. This scoring strategy ensures that if a node matches both benign and malicious subgraphs, its association with malicious subgraphs takes precedence, thereby reinforcing the focus on malicious pattern interpretation and reducing explanatory noise.

In cases, where a node matches both malicious and benign subgraphs (intersected nodes), the SubMatch explainer prioritizes security by treating these nodes as malicious. Their score is computed solely based on matches with malicious subgraphs:
\[
S(v) = \sum_{\mathcal{Q}_m \in \mathcal{B}_{\text{mal}}} \mathbb{I}(v \in V_{\mathcal{Q}_m}).
\]
The detailed procedure of the node risk scoring is outlined in Algorithm~\ref{alg:SubMatch_Explainer}.
Figure~\ref{fig:SubMatch_Explainer} illustrates the SubMatch explainer process. In this figure, the size of each node reflects the number of matches it has with subgraphs in the query box, where larger circles indicate higher match frequency. Red-colored nodes represent matches with malicious subgraphs, while blue-colored nodes correspond to matches with benign subgraphs. This visualization highlights how the SubMatch explainer provides fine-grained interpretability by linking nodes in the target graph to known behavioral prototypes.

\begin{algorithm}[tb]
\caption{Node Risk Scoring}
\label{alg:SubMatch_Explainer}
\begin{algorithmic}[1]
\State \textbf{Input:} Target graph $\mathcal{T}$, verified query box $\mathcal{B} = \mathcal{B}_{\text{mal}} \cup \mathcal{B}_{\text{ben}}$
\State Initialize node scores: $S(v) \leftarrow 0$ for all $v \in V_{\mathcal{T}}$
\For{each subgraph $\mathcal{Q}_m \in \mathcal{B}_{\text{mal}}$}
    \State $\mathcal{M} \leftarrow$ matched nodes of $\mathcal{Q}_m$ in $\mathcal{T}$
    \For{each node $v \in \mathcal{M}$}
        \State $S(v) \leftarrow S(v) + 1$
    \EndFor
\EndFor
\For{each subgraph $\mathcal{Q}_b \in \mathcal{B}_{\text{ben}}$}
    \State $\mathcal{M} \leftarrow$ matched nodes of $\mathcal{Q}_b$ in $\mathcal{T}$
    \For{each node $v \in \mathcal{M}$}
        \If{$S(v) \leq 0$}
            \State $S(v) \leftarrow S(v) - 1$
        \EndIf
    \EndFor
\EndFor
\State \textbf{Output:} Node score map $S(v)$ for all $v \in V_{\mathcal{T}}$
\end{algorithmic}
\end{algorithm}

\section{Experimental Results}
\label{sec:experiments}
This section presents the experimental evaluation of our proposed framework for malware detection with dual explainability. We conducted experiments to assess the effectiveness of each stage of the framework, from generating CFGs to training the GNN model, extracting and verifying subgraphs, and performing subgraph matching for prediction and deeper explainability. Our analysis focuses on the accuracy of malware detection and the quality of explanations provided by the SubMatch explainer. The results demonstrate the framework's ability to achieve high prediction performance, while providing interpretable insights into the detected malicious patterns.

The experiments in this study were conducted on a diverse collection of datasets to evaluate the effectiveness of the proposed framework. For malicious samples, we extracted 115 samples from the BODMAS~\cite{BODMAS} and PMML~\cite{PMML} datasets. Both datasets are well-known repositories containing a variety of malware samples, providing a comprehensive representation of real-world malicious behaviors. For benign samples, we utilized 320 samples extracted from the DikeDataset~\cite{DikeDataset}, a reliable source of non-malicious programs. 

The preprocessing stage is a critical component of the framework, ensuring that raw data is converted into a format suitable for analysis and training. For this study, the CFG of each sample is dynamically extracted using the Python library angr~\cite{shoshitaishvili2016state, stephens2016driller, shoshitaishvili2015firmalice}. 

To represent node attributes within the CFG, we utilized assembly-level instruction encoding~\cite{Dynamic}. Each instruction is encoded as a fixed 438-bit vector, consisting of distinct fields that capture the structural details of the instruction: Option (5 bits), Prefix (9 bits), Opcode (256 bits), ModRM (20 bits), SIB (20 bits), Displacement (64 bits), and Immediate (64 bits). These fixed-length vectors serve as the inputs to an AE, which generates a compressed embedding for each node. The AE architecture consists of an encoder with two hidden layers containing 256 and 128 neurons, followed by a bottleneck layer with 64 neurons. The decoder mirrors the encoder with two hidden layers of 128 and 256 neurons, reconstructing the original input. The model is trained for 700 epochs with a learning rate of 0.01, ensuring sufficient training to optimize the embeddings. The embeddings extracted from the bottleneck layer provide a 64-dimensional representation for each node, capturing the most salient features of the instructions. 
After preprocessing and embedding, the dataset is split into training and testing subsets. 80\% percent of the data is used for training, and the remaining twenty percent is set aside for testing.

For graph-based operations and deep learning, we employed PyTorch Geometric, a state-of-the-art library tailored for graph-structured data. Our experiments utilized three widely adopted GNN architectures: Graph Convolutional Network (GCN), GraphSAGE, and Graph Attention Network (GAT). Each model consists of three layers with 64 hidden units per layer and incorporates a dropout rate of 50\% to reduce overfitting. The models were trained for 250 epochs using the Adam optimizer with a learning rate of 0.0001 and a weight decay of 0.0005.

To interpret the GNN decisions, we integrated the Captum library through a custom CaptumExplainer module. This module implements three attribution techniques (Integrated Gradients, Guided Backpropagation, and Saliency), which provide insights into the internal decision-making processes of the GNN by identifying critical substructures that contribute most to the prediction outcomes.

\begin{figure*}[h]
    \centering

    % First Row
    \begin{subfigure}[b]{0.48\textwidth}
        \centering
        \includegraphics[width=\linewidth]{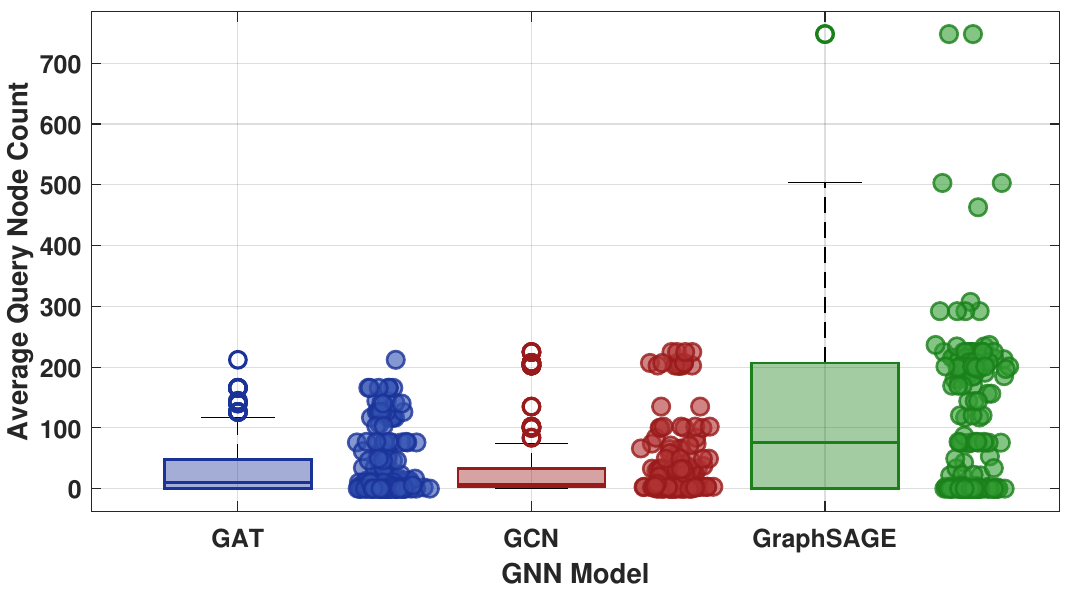}
        \caption{}
        \label{fig:GNN_node}
    \end{subfigure}
    \hfill
    \begin{subfigure}[b]{0.48\textwidth}
        \centering
        \includegraphics[width=\linewidth]{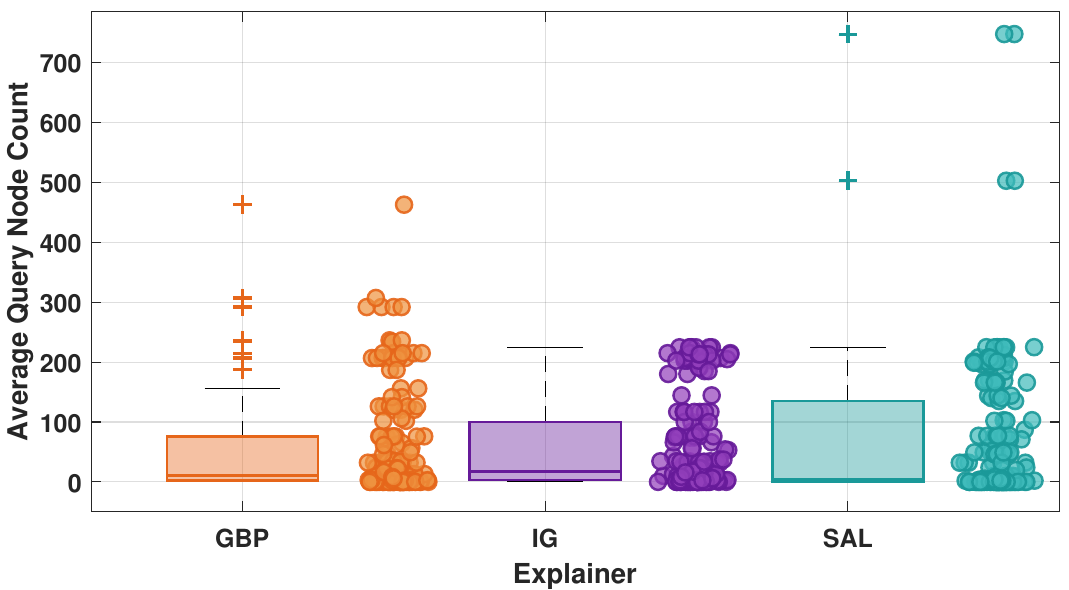}
        \caption{}
        \label{fig:Explainer_node}
    \end{subfigure}

    \vspace{0.2cm}

    % Second Row
    \begin{subfigure}[b]{0.48\textwidth}
        \centering
        \includegraphics[width=\linewidth]{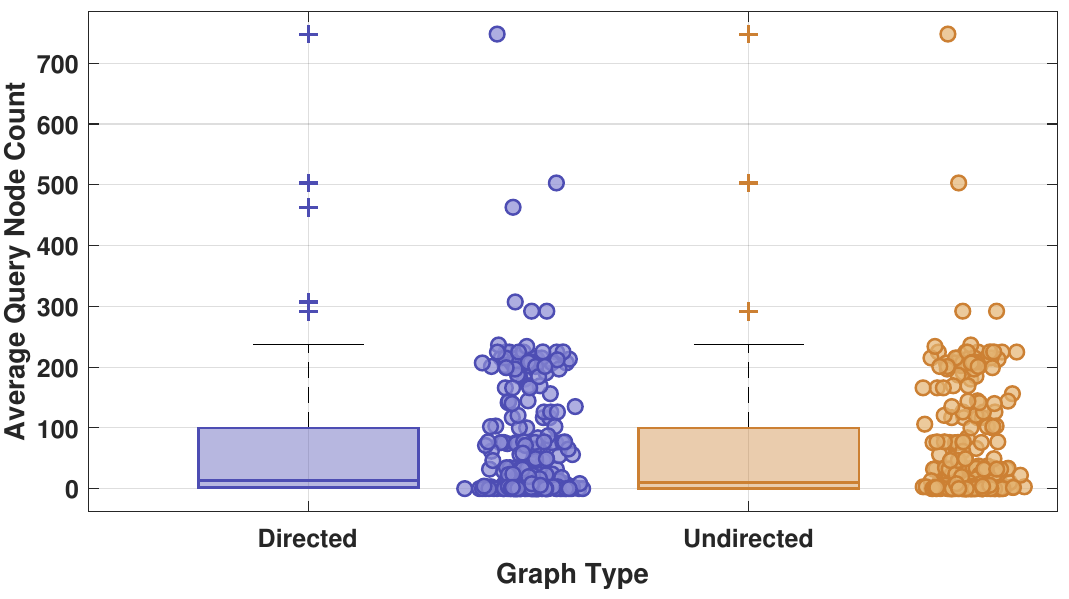}
        \caption{}
        \label{fig:Graph_node}
    \end{subfigure}
    \hfill
    \begin{subfigure}[b]{0.48\textwidth}
        \centering
        \includegraphics[width=\linewidth]{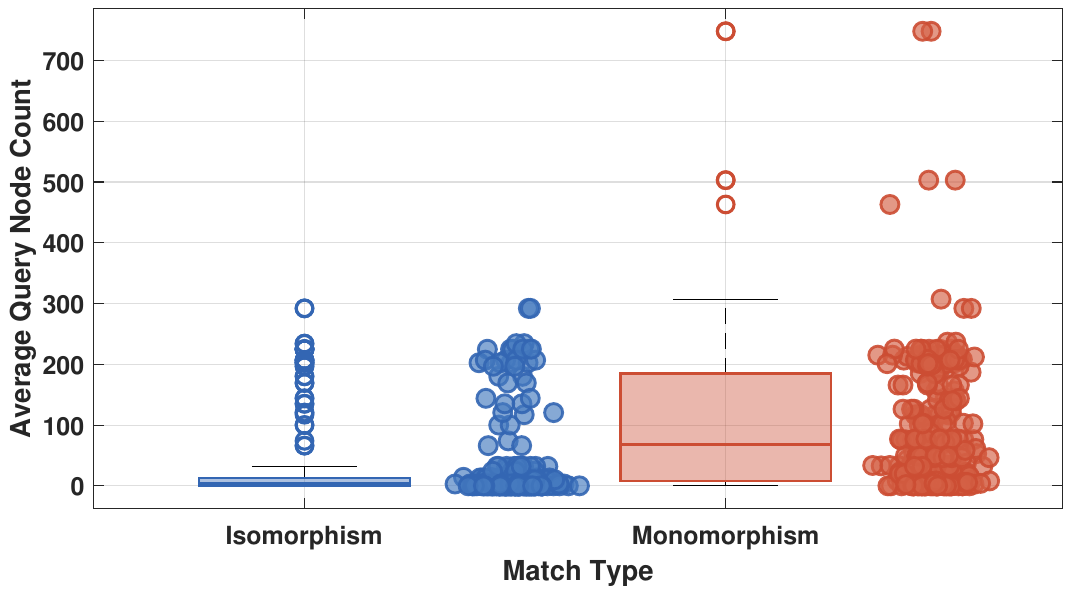}
        \caption{}
        \label{fig:Match_node}
    \end{subfigure}

    \caption{Boxplots of the average number of matched query nodes across different experimental factors under the exact structural and attribute matching setting: (a) GNN model comparison (GraphSAGE, GCN, GAT), (b) explainer comparison (IG, GBP, SAL), (c) graph type comparison (directed vs. undirected), and (d) matching strategy comparison (monomorphism vs. isomorphism).}
    \label{fig:average}
\end{figure*}

For subgraph-level explainability and pattern verification, we adopted the VF2 algorithm, a well-established method for subgraph matching. In our implementation, we support both isomorphism and monomorphism matching to allow for both strict and relaxed pattern detection. Furthermore, all matching experiments were performed on both directed and undirected versions of the CFGs to ensure a comprehensive analysis of graph structure in relation to malicious behavior. The matching and visualization processes were carried out using NetworkX, a versatile Python library for graph analysis.

\begin{figure*}[h]
    \centering

    % First Row
    \begin{subfigure}[b]{0.48\textwidth}
        \centering
        \includegraphics[width=\linewidth]{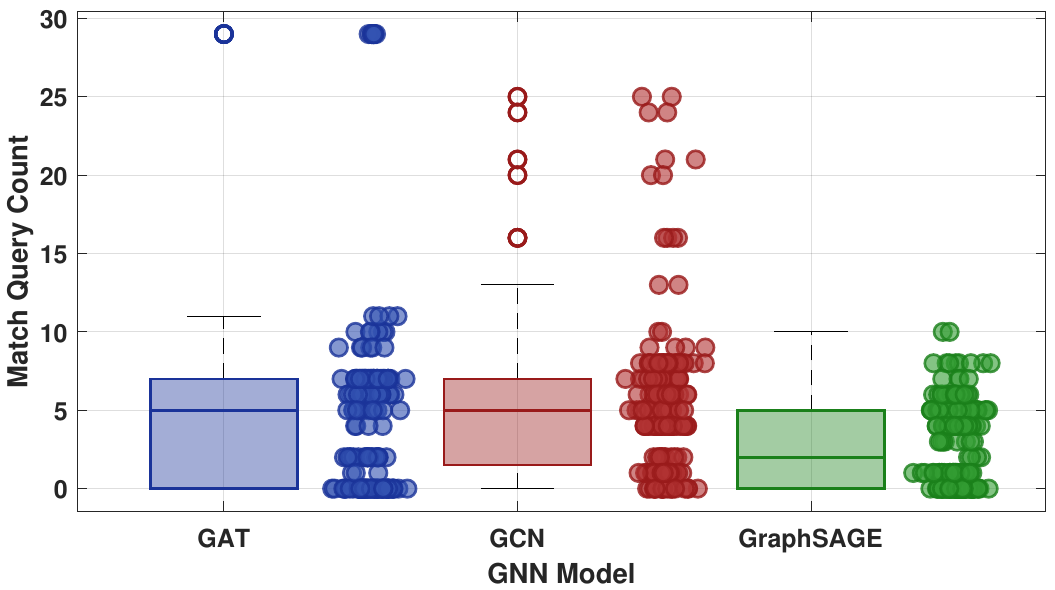}
        \caption{}
        \label{fig:GNN_query}
    \end{subfigure}
    \hfill
    \begin{subfigure}[b]{0.48\textwidth}
        \centering
        \includegraphics[width=\linewidth]{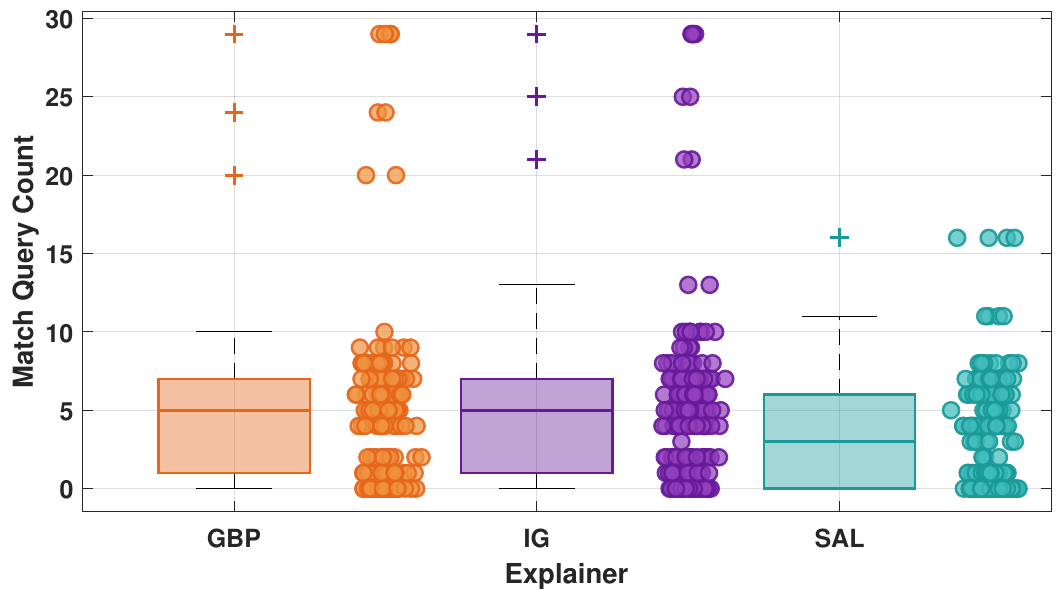}
        \caption{}
        \label{fig:Explainer_query}
    \end{subfigure}

    \vspace{0.2cm}

    % Second Row
    \begin{subfigure}[b]{0.48\textwidth}
        \centering
        \includegraphics[width=\linewidth]{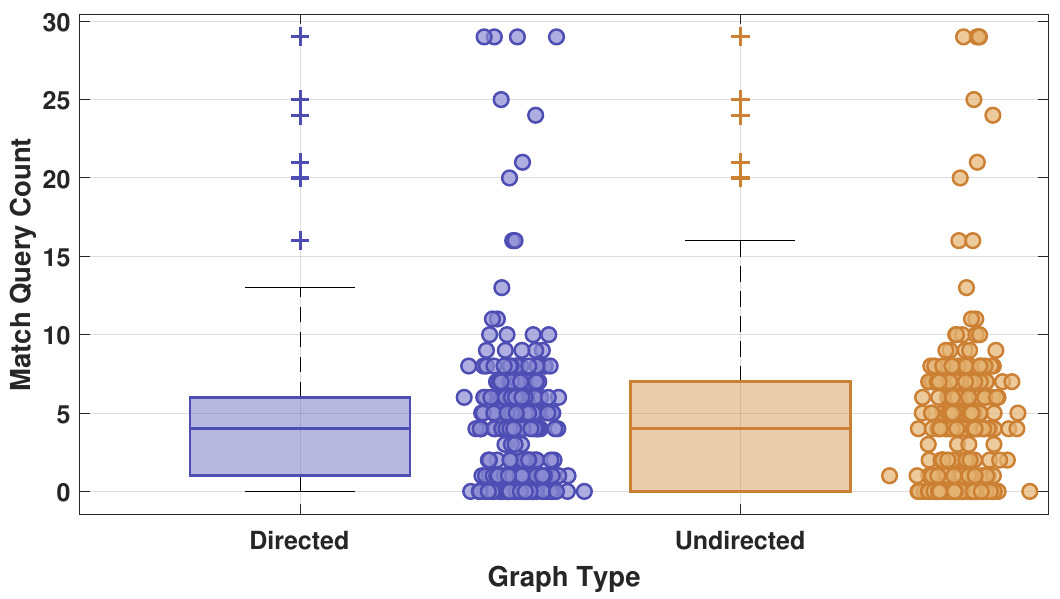}
        \caption{}
        \label{fig:Graph_query}
    \end{subfigure}
    \hfill
    \begin{subfigure}[b]{0.48\textwidth}
        \centering
        \includegraphics[width=\linewidth]{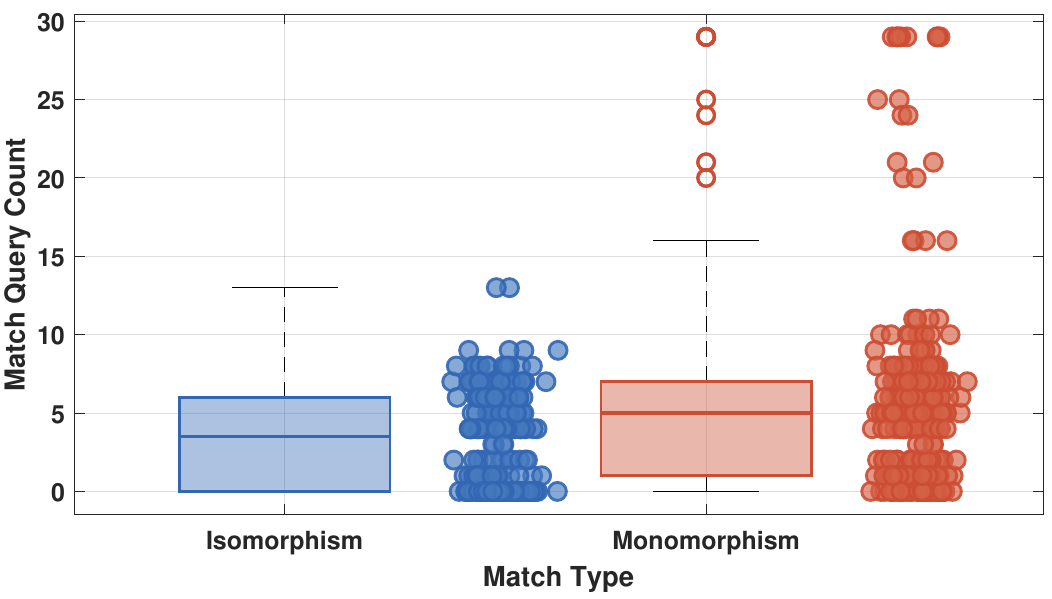}
        \caption{}
        \label{fig:Match_query}
    \end{subfigure}

    \caption{Boxplots of the number of matched queries across different experimental factors under the exact structural and attribute matching setting: (a) GNN model comparison (GraphSAGE, GCN, GAT), (b) explainer comparison (IG, GBP, SAL), (c) graph type comparison (directed vs. undirected), and (d) matching strategy comparison (monomorphism vs. isomorphism).}

    \label{fig:match_count_grid}
\end{figure*}

To ensure the robustness and generalizability of our findings, we performed a comprehensive grid search across various configurations. This included combinations of the three GNN architectures (GCN, GraphSAGE, and GAT), the three attribution methods from Captum, the two subgraph matching types (isomorphism and monomorphism), and both directed and undirected graph representations. This extensive exploration allowed us to systematically evaluate the interaction between model structures, explainability techniques, and matching strategies, ultimately identifying the most effective setup for accurate and interpretable malware detection.

\begin{figure*}[h]
    \centering
    \begin{subfigure}[b]{0.47\linewidth}
        \centering
        \includegraphics[width=\linewidth]{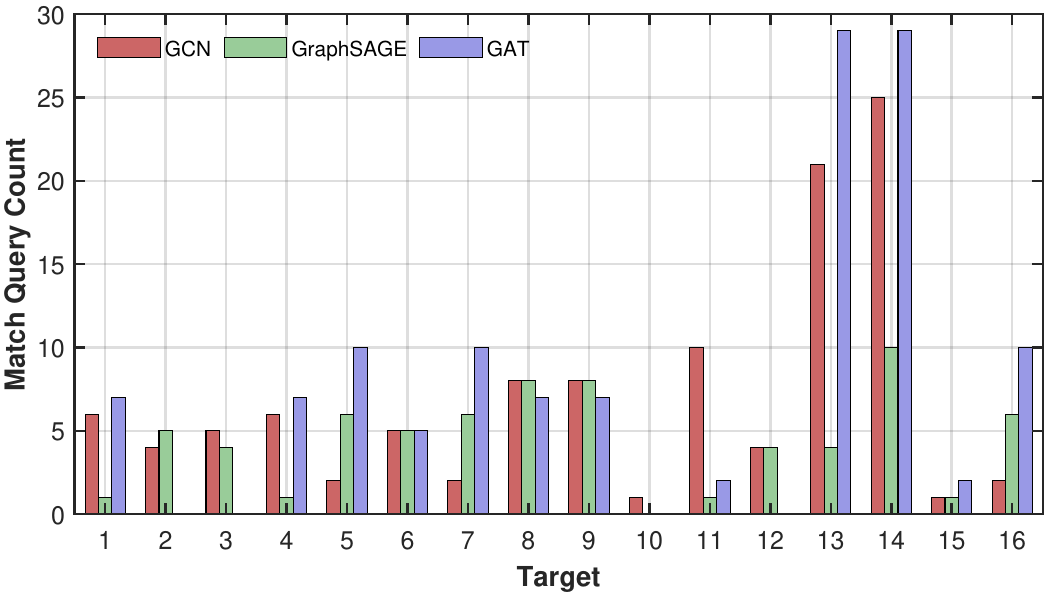}
        \caption{}
        \label{fig:match_count}
    \end{subfigure}
    \hfill
    \begin{subfigure}[b]{0.48\linewidth}
        \centering
        \includegraphics[width=\linewidth]{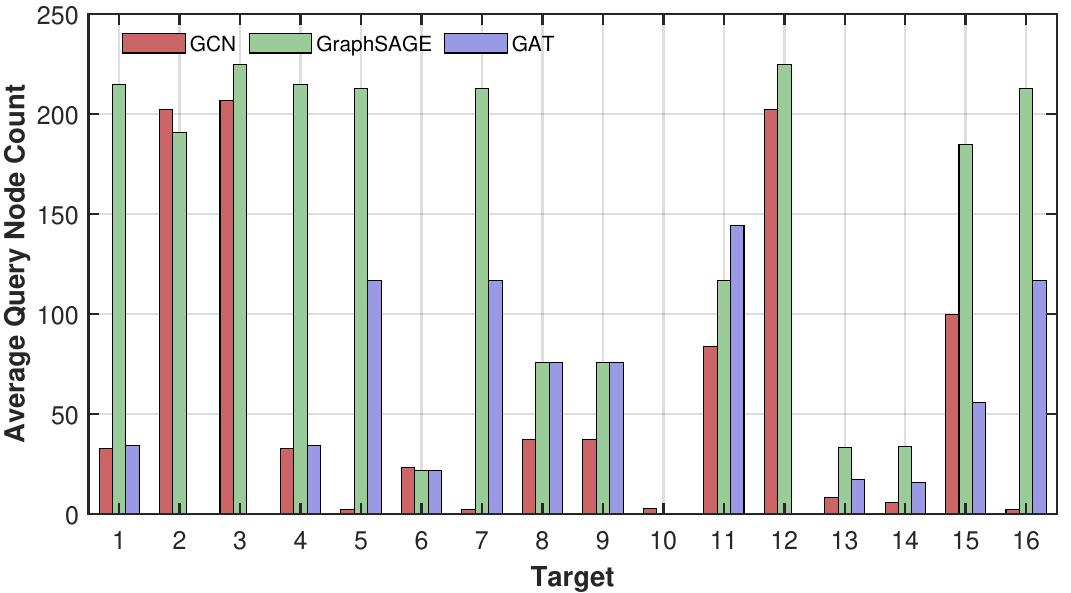}
        \caption{}
        \label{fig:average_node}
    \end{subfigure}

    \caption{Evaluation of SubMatch explainer performance across GNN models under the exact structural and attribute matching setting.}
    \label{fig:Inexact_Explainer}
\end{figure*}

The training phase of the framework achieved high classification performance, with a training accuracy of 91.95\% for both the GCN and GAT models, and 93.10\% for GraphSAGE. Following model training, the correctly labeled samples from the training set were passed through the three explainers (Integrated Gradients, Guided Backpropagation, and Saliency). Each explainer produces a weighted version of the input graph, in which importance scores were assigned to edges based on their contribution to the model’s predictions.

Subsequently, the GEC algorithm was applied to extract the most informative subgraphs from the weighted graphs. Subgraphs that passed the verification process (i.e., those for which the GNN’s prediction remained consistent with the original sample label) were stored in verified query repositories corresponding to their specific configuration. Each unique combination of explainer method (Integrated Gradients, Guided Backpropagation, or Saliency), GNN architecture (GCN, GraphSAGE, or GAT), subgraph matching type (isomorphism or monomorphism), and graph directionality (directed or undirected) was assigned its own query box as part of the broader grid search strategy.

\subsection{Performance Evaluation of Exact Structural and Attribute Matching Approach}
In this set of experiments, we evaluate the performance of the SubMatch explainer under the exact structural and attribute matching setting.To facilitate interpretation, the abbreviations used in the figures are as follows: IG (Integrated Gradients), GBP (Guided Backpropagation), and SAL (Saliency).
\begin{figure}[h]
    \centering
    \includegraphics[width=\linewidth]{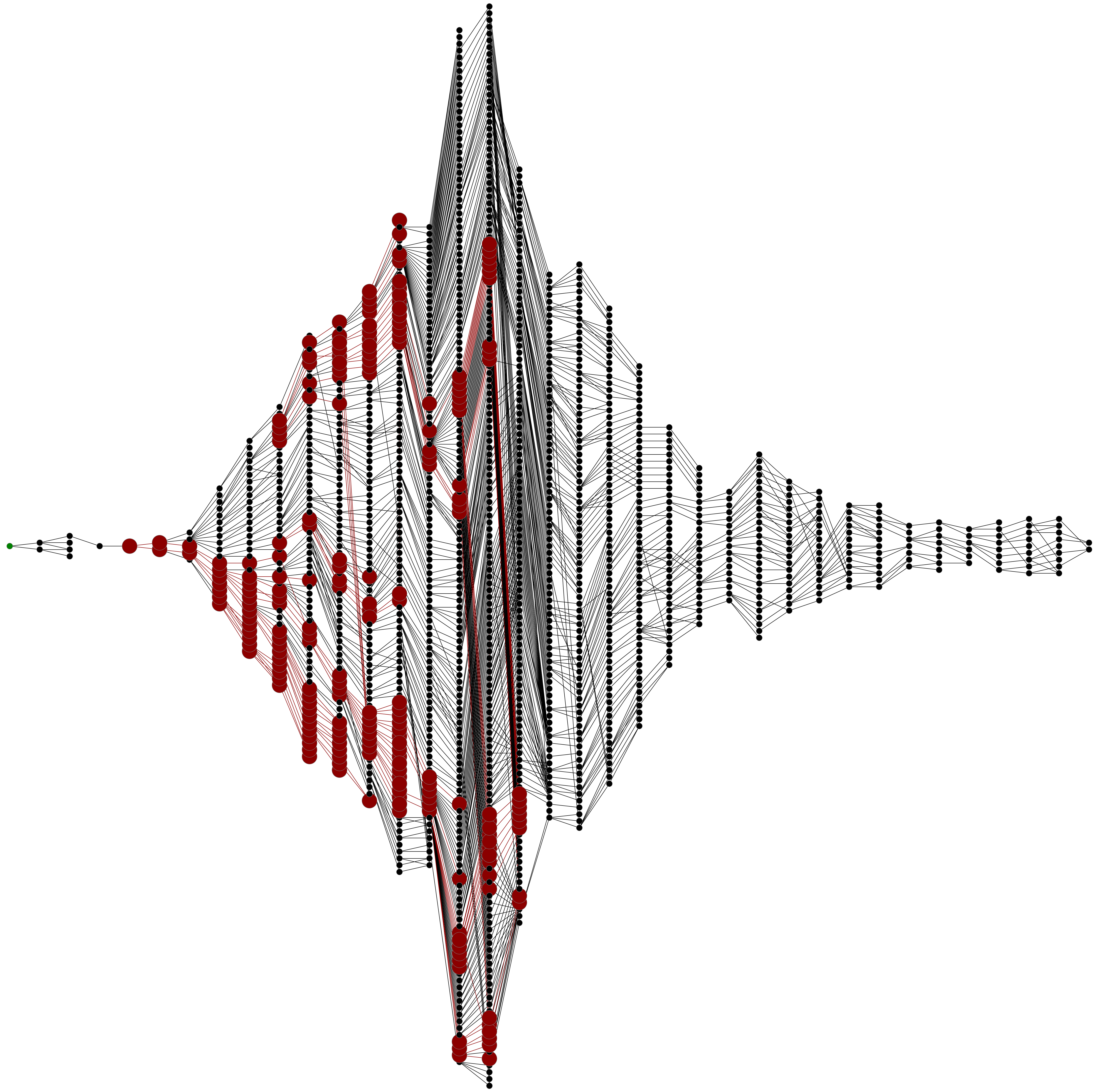}
    \caption{Dual explanation provided by the SubMatch explainer for a representative malicious target under the exact structural and attribute matching approach.}
    \label{fig:Exact_Sample}
\end{figure}

We performed a comprehensive grid search across all possible combinations of three GNN architectures (GCN, GAT, and GraphSAGE), three GNN explainers (Integrated Gradients, Guided Backpropagation, and Saliency), two matching strategies (isomorphism and monomorphism), and two graph types (directed and undirected). Among the 23 malicious test samples, the maximum number of successful matches with verified queries was 16.
The comparative results are illustrated in Figure~\ref{fig:average}, which presents boxplots of the average number of matched query nodes across different factors. Specifically, Figure~\ref{fig:GNN_node} compares the GNN models, Figure~\ref{fig:Explainer_node} compares the explainers, Figure~\ref{fig:Graph_node} compares directed and undirected graphs, and Figure~\ref{fig:Match_node} compares monomorphism and isomorphism matching strategies.

Observations reveal that, in terms of the average number of matched query nodes, GraphSAGE consistently outperforms the other GNN models. Regarding the matching strategy, monomorphism achieves notably better performance compared to isomorphism. In contrast, the choice of graph type (directed vs. undirected) and the choice of explainer (IG, GBP, SAL) exhibit no significant differences with respect to this metric.
Similarly, Figure~\ref{fig:match_count_grid} shows the corresponding boxplots for the number of matched queries (i.e., the number of verified queries in the query box that were successfully matched with target samples). In this case, the results suggest that there is no significant variation across different GNN models, explainers, graph types, or matching methods.

To analyze the performance across GNN models in more detail, we fixed the explainer to Integrated Gradients, the graph type to undirected, and the matching strategy to monomorphism. Under this fixed setting, the SubMatch explainer’s matching results are summarized in Figure~\ref{fig:Inexact_Explainer}. Among the 23 malicious test samples, 16 samples were successfully matched with verified queries using the GCN model, 15 samples with GraphSAGE, and 12 with GAT. A detailed breakdown is provided in Figure~\ref{fig:match_count}, which presents a bar chart indicating, for each sample and GNN model, the number of successful matches with verified subgraphs. Furthermore, Figure~\ref{fig:average_node} reports the average number of nodes from the matched queries.
In terms of node-level correspondence, GraphSAGE achieves a higher average matching quality compared to GCN and GAT, reflecting its superior representational capabilities in this setting. 

Importantly, we observed that none of the malicious target samples were mistakenly matched with benign queries in this exact setting; all matches involved only verified malicious queries, preserving the semantic fidelity of the detection. To demonstrate the interpretability of the method, Figure~\ref{fig:Exact_Sample} visualizes the output of the SubMatch explainer applied to a representative malicious sample. Red circles highlight regions matched with malicious subgraphs from the verified query box, while black circles denote unmatched regions. The size of each circle is proportional to the frequency of matches, with larger circles indicating nodes that matched more frequently with verified queries.

\subsection{Performance Evaluation of Exact Structural Matching with Approximate Attribute Matching Approach}
In this evaluation, we analyze the effectiveness of the SubMatch explainer under the exact structural matching with approximate attribute matching setting. Consistent with earlier findings, the choice of GNN architecture and matching strategy exerts a more substantial influence on matching performance compared to other configuration factors, such as graph directionality (directed vs. undirected) or the choice of first-level GNN explainer.
To maintain consistency, we present the results based on a representative configuration that employs GraphSAGE, treats graphs as undirected, utilizes monomorphism-based matching through the VF2 algorithm, and applies Integrated Gradients for computing node importance scores. To enable approximate attribute matching, we use a cosine similarity threshold of 0.9, allowing nodes with semantically similar embeddings to be considered equivalent even when exact attribute values differ.

\begin{figure}[h]
    \centering
    \includegraphics[width=\linewidth]{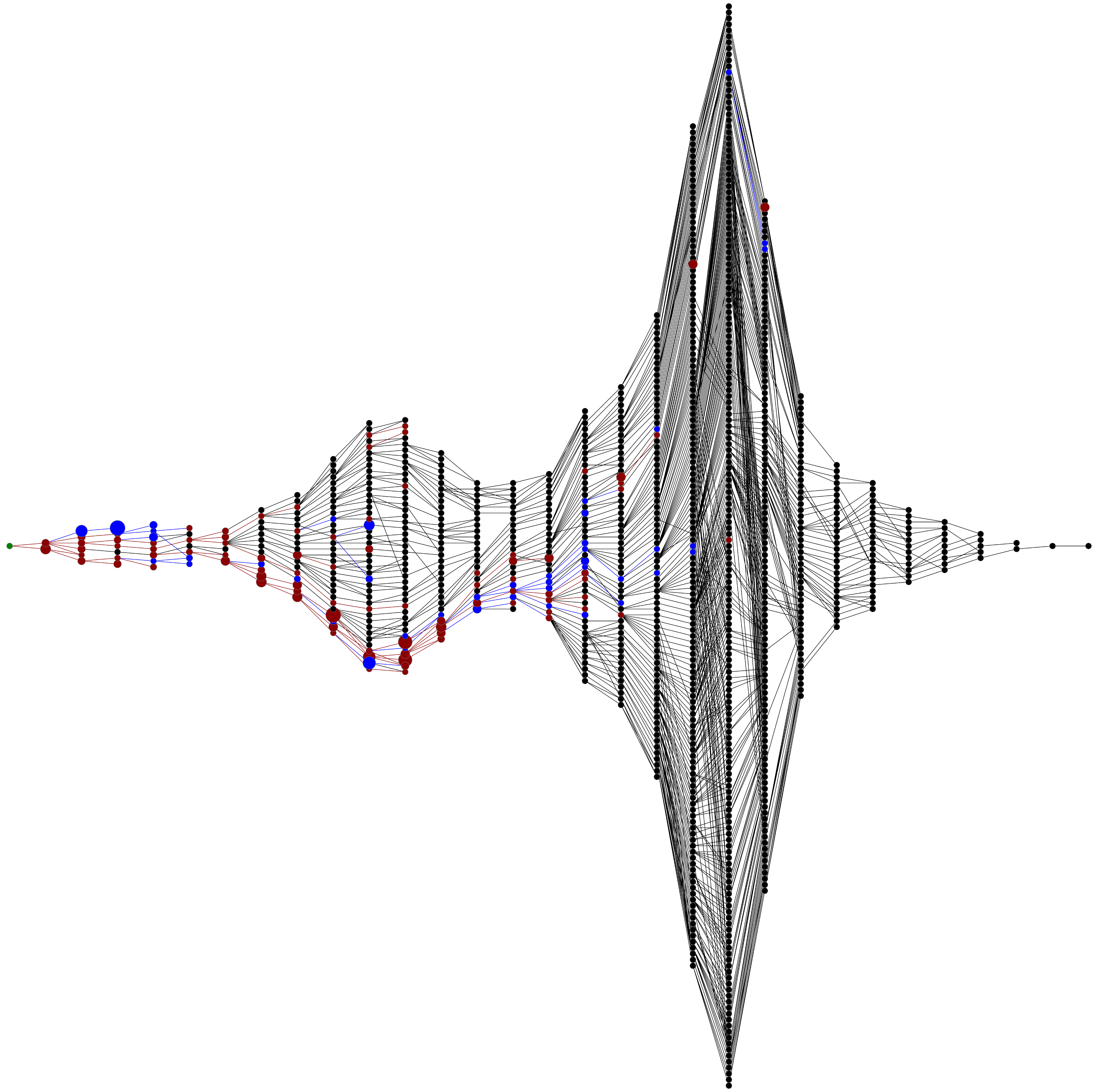}
    \caption{Dual interpretation of a malicious targett sample through SubMatch explainer under the configuration of exact structural matching and relaxed attribute similarity.}
    \label{fig:Inexact}
\end{figure}

Figure~\ref{fig:Inexact} displays the output of the SubMatch explainer under this relaxed matching strategy. In the visualization, red nodes indicate matches with verified malicious queries, while blue nodes correspond to matches with benign queries. Due to the approximate nature of attribute matching, some nodes are matched with both query types, highlighting the nuanced interpretability offered by the SubMatch explainer in capturing overlapping semantic patterns.

\section{\textbf{Conclusion}}
\label{sec:conclusion}
This work presents a novel explainable malware detection framework by integrating CFG extraction, a node feature embedding module, a GNN model, a GNN explainer, and subgraph-level reasoning to provide both high classification accuracy and interpretable insights. The core contribution lies in the design and implementation of the SubMatch explainer, an innovative GNN explainer that enhances model transparency by leveraging verified subgraphs and subgraph matching techniques.
Our methodology integrates dynamic CFG extraction, node-level embedding via autoencoders, GNN-based classification, and dual explainability. By applying both exact and approximate subgraph matching strategies, we introduced a mechanism for attributing model predictions to known and verified subgraph patterns, enabling robust explanations. The SubMatch explainer not only identifies important nodes but also scores them based on their correspondence with known patterns, offering a granular and intuitive view of a model’s decision-making process.
Experimental results across various GNN architectures and explainer configurations demonstrate that GraphSAGE consistently outperforms GCN and GAT in terms of both matching quality and interpretability. In particular, the exact structural and attribute matching setting shows the highest precision, while the approximate attribute matching configuration offered greater flexibility and semantic richness.
In summary, the proposed framework and SubMatch explainer advance the state of explainable malware detection by bridging the gap between model accuracy and interpretability. This framework is particularly valuable for security analysts who require not only accurate predictions but also an understanding of the underlying rationale. Future work may extend this framework by incorporating temporal dynamics from execution traces, exploring contrastive subgraph learning, or integrating online learning to adapt to evolving malware behaviors.

\bibliographystyle{IEEEtran}
\bibliography{ref1}

\begin{thebibliography}{10}
\providecommand{\url}[1]{#1}
\csname url@samestyle\endcsname
\providecommand{\newblock}{\relax}
\providecommand{\bibinfo}[2]{#2}
\providecommand{\BIBentrySTDinterwordspacing}{\spaceskip=0pt\relax}
\providecommand{\BIBentryALTinterwordstretchfactor}{4}
\providecommand{\BIBentryALTinterwordspacing}{\spaceskip=\fontdimen2\font plus
\BIBentryALTinterwordstretchfactor\fontdimen3\font minus \fontdimen4\font\relax}
\providecommand{\BIBforeignlanguage}[2]{{%
\expandafter\ifx\csname l@#1\endcsname\relax
\typeout{** WARNING: IEEEtran.bst: No hyphenation pattern has been}%
\typeout{** loaded for the language `#1'. Using the pattern for}%
\typeout{** the default language instead.}%
\else
\language=\csname l@#1\endcsname
\fi
#2}}
\providecommand{\BIBdecl}{\relax}
\BIBdecl

\bibitem{01}
R.~Sun, S.~Guo, J.~Guo, W.~Li, X.~Zhang, X.~Guo, and Z.~Pan, ``{GraphMoCo}: A graph momentum contrast model for large-scale binary function representation learning,'' \emph{Neurocomputing}, p. 127273, 2024.

\bibitem{02}
H.~Peng, J.~Yang, D.~Zhao, X.~Xu, Y.~Pu, J.~Han, X.~Yang, M.~Zhong, and S.~Ji, ``{MalGNE}: Enhancing the performance and efficiency of cfg-based malware detector by graph node embedding in low dimension space,'' \emph{IEEE Transactions on Information Forensics and Security}, 2024.

\bibitem{03}
M.~T. Nguyen, V.~H. Nguyen, and N.~Shone, ``Using deep graph learning to improve dynamic analysis-based malware detection in {PE} files,'' \emph{Journal of Computer Virology and Hacking Techniques}, vol.~20, no.~1, pp. 153--172, 2024.

\bibitem{04}
T.~Li, Y.~Luo, X.~Wan, Q.~Li, Q.~Liu, R.~Wang, C.~Jia, and Y.~Xiao, ``A malware detection model based on imbalanced heterogeneous graph embeddings,'' \emph{Expert Systems with Applications}, vol. 246, p. 123109, 2024.

\bibitem{05}
C.~Liu, B.~Li, J.~Zhao, W.~Feng, X.~Liu, and C.~Li, ``{A2-CLM}: Few-shot malware detection based on adversarial heterogeneous graph augmentation,'' \emph{IEEE Transactions on Information Forensics and Security}, vol.~19, pp. 2023--2038, 2024.

\bibitem{06}
S.~Chen, B.~Lang, H.~Liu, Y.~Chen, and Y.~Song, ``Android malware detection method based on graph attention networks and deep fusion of multimodal features,'' \emph{Expert Systems with Applications}, vol. 237, p. 121617, 2024.

\bibitem{07}
P.~Feng, L.~Gai, L.~Yang, Q.~Wang, T.~Li, N.~Xi, and J.~Ma, ``{DawnGNN}: Documentation augmented {Windows} malware detection using graph neural network,'' \emph{Computers \& Security}, vol. 140, p. 103788, 2024.

\bibitem{08}
Y.~Hei, R.~Yang, H.~Peng, L.~Wang, X.~Xu, J.~Liu, H.~Liu, J.~Xu, and L.~Sun, ``{HAWK}: Rapid {Android} malware detection through heterogeneous graph attention networks,'' \emph{IEEE Transactions on Neural Networks and Learning Systems}, vol.~35, no.~4, pp. 4703--4717, 2024.

\bibitem{09}
L.~Deng, H.~Wen, M.~Xin, H.~Li, Z.~Pan, and L.~Sun, ``Enimanal: Augmented cross-architecture {IoT} malware analysis using graph neural networks,'' \emph{Computers \& Security}, vol. 132, p. 103323, 2023.

\bibitem{10}
Z.~Wang, K.~Zeng, J.~Wang, and D.~Li, ``{FAGnet}: Family-aware-based {Android} malware analysis using graph neural network,'' \emph{Knowledge-Based Systems}, vol. 289, p. 111531, 2024.

\bibitem{11}
J.~Gu, H.~Zhu, Z.~Han, X.~Li, and J.~Zhao, ``{GSEDroid}: {GNN-based} {Android} malware detection framework using lightweight semantic embedding,'' \emph{Computers \& Security}, vol. 140, p. 103807, 2024.

\bibitem{12}
R.~Yumlembam, B.~Issac, S.~M. Jacob, and L.~Yang, ``{IoT}-based {Android} malware detection using graph neural network with adversarial defense,'' \emph{IEEE Internet of Things Journal}, vol.~10, no.~10, pp. 8432--8444, 2023.

\bibitem{13}
X.~Ling, L.~Wu, S.~Wang, T.~Ma, F.~Xu, A.~X. Liu, C.~Wu, and S.~Ji, ``Multilevel graph matching networks for deep graph similarity learning,'' \emph{IEEE Transactions on Neural Networks and Learning Systems}, vol.~34, no.~2, pp. 799--813, 2023.

\bibitem{14}
F.~Ou and J.~Xu, ``{S3Feature}: A static sensitive subgraph-based feature for {Android} malware detection,'' \emph{Computers \& Security}, vol. 112, p. 102513, 2022.

\bibitem{ZHEN2025110524}
Y.~Zhen, D.~Tian, X.~Fu, and C.~Hu, ``A novel malware detection method based on audit logs and graph neural network,'' \emph{Engineering Applications of Artificial Intelligence}, vol. 152, p. 110524, 2025.

\bibitem{Our_Survey}
H.~Shokouhinejad, R.~Razavi-Far, H.~Mohammadian, M.~Rabbani, S.~Ansong, G.~Higgins, and A.~A. Ghorbani, ``Recent advances in malware detection: Graph learning and explainability,'' \emph{arXiv preprint arXiv:2502.10556}, 2025.

\bibitem{Hesam}
H.~Mohammadian, G.~Higgins, S.~Ansong, R.~Razavi-Far, and A.~A. Ghorbani, ``Explainable malware detection through integrated graph reduction and learning techniques,'' \emph{arXiv preprint arXiv:2412.03634}, 2024.

\bibitem{10.5555/3454287.3455116}
R.~Ying, D.~Bourgeois, J.~You, M.~Zitnik, and J.~Leskovec, \emph{GNNExplainer: generating explanations for graph neural networks}, 2019.

\bibitem{PG}
D.~Luo, W.~Cheng, D.~Xu, W.~Yu, B.~Zong, H.~Chen, and X.~Zhang, ``Parameterized explainer for graph neural network,'' ser. NIPS '20, 2020.

\bibitem{captum}
N.~Kokhlikyan, V.~Miglani, M.~Martin, E.~Wang, B.~Alsallakh, J.~Reynolds, A.~Melnikov, N.~Kliushkina, C.~Araya, S.~Yan, and O.~Reblitz-Richardson, ``Captum: A unified and generic model interpretability library for pytorch,'' \emph{arXiv preprint arXiv:2009.07896}, 2020.

\bibitem{subgraphx}
H.~Yuan, H.~Yu, J.~Wang, K.~Li, and S.~Ji, ``On explainability of graph neural networks via subgraph explorations,'' in \emph{Proceedings of the 38th International Conference on Machine Learning}, ser. Proceedings of Machine Learning Research, M.~Meila and T.~Zhang, Eds., vol. 139, 18--24 Jul 2021, pp. 12\,241--12\,252.

\bibitem{8293854}
M.~Fan, J.~Liu, X.~Luo, K.~Chen, Z.~Tian, Q.~Zheng, and T.~Liu, ``Android malware familial classification and representative sample selection via frequent subgraph analysis,'' \emph{IEEE Transactions on Information Forensics and Security}, vol.~13, no.~8, pp. 1890--1905, 2018.

\bibitem{LU2024123922}
X.~Lu, J.~Zhao, S.~Zhu, and P.~Lio, ``Sndgcn: Robust android malware detection based on subgraph network and denoising gcn network,'' \emph{Expert Systems with Applications}, vol. 250, p. 123922, 2024.

\bibitem{ALAM2015212}
S.~Alam, R.~Horspool, I.~Traore, and I.~Sogukpinar, ``A framework for metamorphic malware analysis and real-time detection,'' \emph{Computers \& Security}, vol.~48, pp. 212--233, 2015.

\bibitem{NCP}
H.~Shokouhinejad, R.~Razavi-Far, G.~Higgins, and A.~A. Ghorbani, ``{Node-Centric Pruning}: A novel graph reduction approach,'' \emph{Machine Learning and Knowledge Extraction}, vol.~6, no.~4, pp. 2722--2737, 2024.

\bibitem{Dynamic}
H.~Shokouhinejad, G.~Higgins, R.~Razavi-Far, H.~Mohammadian, and A.~A. Ghorbani, ``On the consistency of gnn explanations for malware detection,'' \emph{arXiv preprint arXiv:2504.16316}, 2025.

\bibitem{BODMAS}
L.~Yang, A.~Ciptadi, I.~Laziuk, A.~Ahmadzadeh, and G.~Wang, ``Bodmas: An open dataset for learning based temporal analysis of pe malware,'' in \emph{2021 IEEE Security and Privacy Workshops (SPW)}.\hskip 1em plus 0.5em minus 0.4em\relax IEEE, 2021, pp. 78--84.

\bibitem{PMML}
{Practical Security Analytics LLC}, ``Pe malware machine learning dataset,'' \url{https://practicalsecurityanalytics.com/pe-malware-machine-learning-dataset/}, 2024, accessed: 2024-08-06.

\bibitem{DikeDataset}
G.-A. Iosif, ``Dikedataset,'' \url{https://github.com/iosifache/DikeDataset}, 2021, accessed on February 27, 2024.

\bibitem{shoshitaishvili2016state}
Y.~Shoshitaishvili, R.~Wang, C.~Salls, N.~Stephens, M.~Polino, A.~Dutcher, J.~Grosen, S.~Feng, C.~Hauser, C.~Kruegel, and G.~Vigna, ``Sok: (state of) the art of war: Offensive techniques in binary analysis,'' 2016.

\bibitem{stephens2016driller}
N.~Stephens, J.~Grosen, C.~Salls, A.~Dutcher, R.~Wang, J.~Corbetta, Y.~Shoshitaishvili, C.~Kruegel, and G.~Vigna, ``Driller: Augmenting fuzzing through selective symbolic execution,'' 2016.

\bibitem{shoshitaishvili2015firmalice}
Y.~Shoshitaishvili, R.~Wang, C.~Hauser, C.~Kruegel, and G.~Vigna, ``Firmalice - automatic detection of authentication bypass vulnerabilities in binary firmware,'' 2015.

\end{thebibliography}

\end{document}